# Dynamic Cybersickness Mitigation via Adaptive FFR and FoV adjustments

Ananth N. Ramaseri-Chandra1,2 and Hassan Reza1

1 SEECS, University of North Dakota, Grand Forks, ND 58203, USA

2 Turtle Mountain College, Belcourt, ND 58367, USA

ananthnag.ramaserich@und.edu , hassan.reza@und.edu

**Abstract.** This paper presents a novel adaptive Virtual Reality (VR) system that aims to mitigate cybersickness, specifically visually induced motion sickness (VIMS) in immersive environments through dynamic, real-time adjustments. The system predicts cybersickness levels in real time using a machine learning (ML) model trained on head tracking and kinematic data. The adaptive system adjusts foveated rendering (FFR) strength and field of view (FOV) to enhance user comfort. With a goal to balance usability with system performance, we believe our approach will optimize both user experience and performance. By adapting responsively to user needs, our work explores the potential of a machine learning-based feedback loop for user experience management, contributing to a user-centric VR system design.

1 Introduction

Virtual Reality is emerging as a cutting-edge technology transforming user interaction in digital environments [1]. Recent trends indicate a remarkable surge in VR adoption, projected to escalate by 15% by 2030 [2]. Cybersickness, a condition similar to motion sickness, is a common experience for up to 80% of first-time VR users [3]. It is a significant usability issue in VR environments [4]. Symptoms include nausea, dizziness, disorientation, and headaches. Visually induced motion sickness (VIMS), a more specific type of cybersickness, can severely impact the user experience and limit the adoption of VR technologies. Mitigating the impact of cybersickness is crucial for a more inclusive VR experience [5]. Over the years, various approaches have been proposed to mitigate cybersickness in VR. These interconnected approaches have been used synergistically to enhance user comfort and extend the duration of VR usage. Some solutions focus on optimizing the VR display through high refresh rates, minimal motionto-photon latency, and field of view restriction [6]. Static approaches apply fixed adjustments that remain constant throughout the VR experience. Some other static methods include adding a virtual nose and snap turning to minimize rotational movements. In contrast, dynamic approaches use real-time data from built-in or external sensors, and they compute and respond in real time to provide more personalized user experiences, putting the user's comfort at the forefront. Foveated rendering (FFR) [6], dynamic field of view (FoV), adjustments, and use of adaptive methods [7] offer significant benefits by tailored cybersickness mitigation to individual users.

Current mitigation techniques do not constantly adapt to variations in individual user responses or changing VR dynamics. An adaptive system that leverages real-time data such as head tracking, physiological feedback, or kinematic metrics can make targeted adjustments based on immediate feedback about the user's cybersickness levels. Furthermore, a real-time feedback loop enables the system to balance comfort with performance, optimizing VR parameters without compromising immersion or creating latency issues. The lack of adaptive VR systems that can predict and mitigate cybersickness in real-time without relying on additional hardware presents a significant challenge. However, the potential impact of a system that can utilize readily available data from VR headsets, such as head tracking and kinematic data, to predict cybersickness likelihood and adjust VR parameters dynamically is profound.

This system, once developed, will address the limitations of current methods by providing personalized mitigation strategies that enhance user comfort without compromising immersion or system performance.

- In this paper, we develop a dynamic system that uses a machine learning model to predict cybersickness likelihood based on continuous real-time data and
- Implement an adaptive system that makes dynamic real-time adjustments to the FFR strength and FOV based on the user's current symptoms and system performance.

We anticipate this approach reduces the computational load associated with VR rendering while simultaneously enhancing the user experience by preemptively adjusting to potential discomfort. Such a robust adaptive system can accommodate various VR scenarios and user profiles, ensuring scalability and inclusivity by accounting for diverse user sensitivities to cybersickness. Our paper starts with an introduction, followed by the related work in Section II and details of our research methodology in Section III. As this is ongoing research, Section IV briefly mentions the anticipated results, and Section V discusses the limitations and future work.

## 2 Related Work

### 2.1 Cybersickness Mitigation Techniques

Current literature on mitigating cybersickness uses a range of static and dynamic approaches. Some foundational techniques, such as FOV restriction and foveated rendering, are used to minimize discomfort and improve user experience in VR. Ang et al. [8] review the existing mitigation techniques and state that foveated Dynamic Cybersickness Mitigation via Adaptive FFR and FoV adjustments 3 rendering reduces computational load by rendering high-resolution details only at the center of the user's gaze. At the same time, peripheral areas remain in lower resolution. FOV restriction limits peripheral vision during motion in a VR scene [9, 10]. Other mitigation techniques include snap turning [9], virtual nose technique [10], dynamic FOV restriction, and EEG-based adjustments [7]. Cybersickness mitigation approaches can be broadly categorized into static and dynamic methods. Static approaches apply fixed mitigations that remain constant regardless of user feedback. Overall, the methods that were dynamic and adaptive were more successful in mitigating cybersickness [8]. Uyan et al. and McGill et al. [7, 11] highlight how dynamic approaches adapt to real-time indicators of cybersickness to adjust VR parameters like navigation speed or visual complexity based on user biofeedback.

### 2.2 Machine learning, Head Tracking and Kinematic data in VR

Machine learning (ML) based adaptive VR systems significantly enhance user experience by personalizing interactions and optimizing system performance. These systems leverage advanced algorithms to adapt to user preferences and real-time data, creating immersive environments that cater to individual needs [12–14].

Deep learning and reinforcement learning enable VR systems to learn from user interactions, improving engagement and immersion over time [15]. Systems like SmartSimVR utilize user-specific data to adjust the virtual environment dynamically, addressing issues like cybersickness and enhancing comfort [16].

Islam et al. feature the use of head-tracking data with machine learning (ML), showing significant promise in predicting cybersickness in VR [17]. Kinematic metrics like head movement position, velocity, acceleration, and jerk can be used to train predictive models [18]. Studies employing such techniques have achieved prediction accuracies from 70% to 93%. However, the potential of this approach is significantly enhanced when combined with multimodal datasets, underscoring the importance of interdisciplinary collaboration. Despite the challenges posed by high computational demands and the risk of overfitting, particularly in real-time applications where VR hardware may not meet the needs of complex models, the potential of this approach is undeniable. It is important to note that models trained on specific VR interactions, such as passive exposure, may face difficulties in generalizing across a wide range of VR experiences.

**2.3 Limitations of existing work**

In this research, we highlight several significant drawbacks that need to be addressed to make VR systems more practical and user-friendly. These drawbacks include the dependence on additional hardware, such as EEG sensors, for monitoring physiological signals and providing feedback. Additionally, latency issues can negatively impact the VR experience and the specificity of symptoms—where systems may only address particular symptoms and not the full spectrum of cybersickness. Static methods do not adapt to individual user differences or changing states during VR interaction, lacking personalization. Current VR systems primarily rely on post-experience evaluations that fail to address discomfort in real time. Such a one-size-fits-all approach results in inconsistent user experiences, where many users continue to experience discomfort and diminished immersion despite existing efforts.

Further research is crucial to refining adaptive methods that use minimal hardware and consider the diverse symptoms of cybersickness. While these adaptive systems offer significant benefits, they also present challenges, such as dependence on accurate user states, challenges in cybersickness detection, implementation complexity, issues with generalizability, and variability in user experiences.

By leveraging continuous data inputs such as head tracking, velocity, acceleration, and user responses, an adaptive VR system can make instantaneous adjustments to key parameters like Field of View (FOV), Foveated Rendering (FFR) strength, and framerate [7, 10, 11]. This real-time adaptability not only anticipates factors contributing to cybersickness but also ensures an optimal balance between system performance and user experience by dynamic adjustments.

**3 Methodology**

The main goal is to optimize the VR user experience by dynamically adjusting system parameters based on real-time cybersickness predictions. The system employs a random forest (RF) [19] machine learning model for cybersickness prediction and adapts to the VR environment using foveated rendering and field of view adjustments. We used an Oculus Quest 2 VR headset with a 72 Hz refresh rate and a high-performance laptop equipped with an NVIDIA 4090 GPU, Intel Core i9 CPU, and 32 GB RAM. The VR scene was developed in Unity Engine 2022LTS [20], integrating Oculus features. Python 3.12 [21] was used for machine learning with sci-kit-learn. Custom C# scripts in Unity handle communication and data acquisition, employing asynchronous methods for data streaming to prevent rendering delays. The random forest model is optimized to an onnx [22, 23] format for Unity, ensuring compatibility and efficient resource management.

## 3.1 System Design

The Dynamic Adjustment System is structured to integrate seamlessly with existing VR setups, providing real-time adaptability without compromising system performance. Our system architecture consists of three main components:

1. Data Acquisition Module: Collects real-time head tracking data (HTD) from the VR headset's sensors.

2. Cybersickness Prediction Module: Utilizes a machine learning model to predict the likelihood of cybersickness based on processed HTD.

3. Adaptive Adjustment and performance module: Implements foveated rendering and FOV adjustments based on predictions to mitigate cybersickness. It also monitors system metrics to ensure adjustments do not adversely affect performance.

## 3.2 Machine Learning Model for Cybersickness Prediction

We used a Random Forest Regressor for the ML model to predict cybersickness in real time. It was chosen for its accuracy and robustness in handling complex relationships of nonlinear data. HTD was collected from 14 participants interacting with a VR environment, and kinematic metrics such as velocity, acceleration, jerk angular velocity, angular acceleration, and angular jerk were computed. Cybersickness scores were calculated for each participant using the Virtual Reality Sickness Questionnaire (VRSQ) [24] and were aligned with the HTD to form a training dataset. The data processing pipeline included standardization, data splitting, and model evaluation. The model was fine-tuned using hyperparameter optimization through GridSearchCV [25]. The model was evaluated with regression metrics [26], yielding a Mean Squared Error (MSE) of 5.0277, Root Mean Squared Error (RMSE) of 2.2422, a Mean Absolute Error (MAE) of 0.6324, and an $R^2$ score of 0.9742 indicating that it could predict cybersickness scores with high accuracyThe model's high $R^2$ score demonstrates its strong capability to explain the variance in cybersickness scores, confirming its effectiveness for real-time prediction in VR environments.

## 3.3 Adaptive adjustments

The Adaptive Adjustment Module aims to reduce cybersickness symptoms in VR environments while maintaining optimal performance. It dynamically adjusts VR scene parameters (Foveated Rendering (FFR) and Field of View (FOV)) using a rule-based approach that evaluates a cybersickness Score from the ML model and framerate against predefined thresholds.

Figure 1 illustrates the flow of the Dynamic Adjustment System, which aims to optimize the user experience in virtual Reality (VR) by dynamically adjusting system parameters based on the predicted likelihood of cybersickness and system performance metrics, particularly the framerate. When the cybersickness score exceeds a set limit and the framerate is below the threshold, the system first increases the strength of FFR and reassesses the score. If the score remains high and the framerate is acceptable, then the FOV is reduced. This iterative feedback loop continues until the cybersickness symptoms are alleviated or the minimum limits for FFR and FOV adjustments are reached while continuously monitoring performance. The adaptive VR system we propose is grounded in control theory and human-computer interaction principles. It utilizes a closed-loop feedback mechanism to predict and mitigate cybersickness in real time, ensuring an optimal user experience without compromising performance.

## 4 Expected Results and Impact

We anticipate that the adaptive session will result in: Dynamic Cybersickness Mitigation via Adaptive FFR and FoV adjustments 7

1. Reduced Cybersickness: Lower VRSQ and SSQ scores compared to the Baseline Session.

2. Maintained or Enhanced Immersion: Similar or improved Presence Questionnaire scores, indicating that adjustments did not detract from the VR experience.

3. Stable System Performance: Framerate and latency metrics remain within acceptable thresholds, demonstrating that the adaptive adjustments do not negatively impact performance.

## 5 Future Work, Scalability and Limitations

This study presents a Dynamic Adjustment System aimed at enhancing the VR user experience by mitigating cybersickness through real-time parameter adjustments. The machine learning model developed for predicting cybersickness has exhibited a high level of predictive accuracy ($R^2$ = 0.9742) utilizing a Random Forest Regressor. However, the integration of this model into the broader adaptive module remains a work in progress. The model's capacity to forecast cybersickness based on head tracking and kinematic data substantiates its reliability as a real-time detection tool, thereby establishing a solid foundation for adaptive VR applications.

Future efforts will focus on the validation of the Dynamic Adjustment Module, a key component designed to dynamically adjust VR parameters such as the intensity of FFR and the FOV based on cybersickness predictions. It is imperative to conduct comprehensive testing across a range of VR environments to ensure that these adjustments effectively alleviate cybersickness while maintaining critical performance metrics, such as frame rate stability and latency. Equally important are user studies conducted across different VR scenarios, which will provide valuable insights into the impact of these adjustments on user comfort, immersion, and overall system responsiveness.

Despite its promising features, the system's scalability may encounter several challenges. Although the Random Forest model is accurate, it needs considerable computational resources, which may restrict its real-time applicability on VR headsets with less processing power. Additional optimization of the model or the investigation of alternative, lighter machine-learning methods may enhance its compatibility with lower-resource devices. Additionally, the diverse configurations of VR applications and hardware necessitate adaptive solutions that can be scaled and tailored to the specific capabilities of various platforms.

Furthermore, the need for recalibration may arise when new VR environments or user demographics are introduced. The model's accuracy may vary based on demographic characteristics or individual user differences, underscoring the importance of adaptive recalibration procedures to accommodate such variability. Currently, the system has been validated using a single dataset; therefore, expanding the dataset to cover a broader range of users and VR scenarios is essential for generalizing its effectiveness. By addressing these limitations and improving the system's scalability, this research aims to establish a robust, adapt- able VR solution that optimizes both user comfort and performance in real time, thereby contributing to more immersive and user-centered VR experiences.

**6 Conclusion**

In conclusion, our study introduces a novel adaptive system that utilizes machine learning for real-time cybersickness prediction in VR environments. Post prediction, dynamic adjustments to Field of View (FOV) and Foveated Rendering (FFR) are made. The goal is to prove that the system shows significant potential to enhance VR comfort without compromising performance or immersion. For the initial part, our Random Forest Regressor model has demonstrated exceptional predictive accuracy, confirming the feasibility of real-time prediction and intervention in VR scenarios. While this approach is promising, testing the system is necessary to show its efficacy and scalability. As VR technology becomes more prevalent, adaptive systems will improve usability and accessibility, making VR experiences more inclusive and enjoyable for all users.

**References**

1. Sushmitha, J., Disha, J.: A survey on virtual reality for medical applications. Journal of Innovative Image Processing (2023)

2. Research, G.V.: Virtual Reality Market Size & Share Report, 2022-

2030 (2020), https://www.grandviewresearch.com/industry-analysis/

virtual-reality-vr-market, accessed 2024-03-12.

3. Nesbitt, K., Nalivaiko, E.: Cybersickness, pp. 1–6. Springer International Publishing, Cham (2018)

4. Khenak, N., Bach, C., Buratto, F.: Understanding the relationship between cybersickness and usability through seven human factors dimensions: An exploratory

comparison of two virtual reality training applications. In: Proceedings of the 18th

"Ergonomie et Informatique Avancée" Conference. p. 1–10. ACM, Bidart France

(Oct 2023)

5. Garrido, L.E., Frías-Hiciano, M., Moreno-Jiménez, M., Cruz, G.N., García-Batista,

Z.E., Guerra-Peña, K., Medrano, L.: Focusing on cybersickness: pervasiveness, latent trajectories, susceptibility, and effects on the virtual reality experience. Virtual

Reality 26, 1347–1371 (2021)

6. Groth: Cybersickness reduction via gaze-contingent image deformation. ACM

Trans. Graph. 43(4) (Jul 2024)

7. Uyan, U., Celikcan, U.: Cdms: A real-time system for eeg-guided cybersickness

mitigation through adaptive adjustment of vr content factors. Displays 83, 102704

(2024)

8. Ang, S., Quarles, J.: Reduction of cybersickness in head mounted displays use: A systematic review and taxonomy of current strategies. Frontiers in Virtual Reality 4, 1027552 (2023)

9. Kelly, J.W., Doty, T.A., Gilbert, S.B., Dorneich, M.C.: Field of view restriction and snap turning as cybersickness mitigation tools. IEEE Transactions on Visualization and Computer Graphics p. 1–9 (2024)

Dynamic Cybersickness Mitigation via Adaptive FFR and FoV adjustments 9

10. Oka, A.A.: Mitigating Vr Cybersickness Caused by Continuous Joystick Movement. Master's thesis, Purdue University (2023)

11. Li, G., McGill, M., Brewster, S., Pollick, F.: A review of electrostimulation-based cybersickness mitigations. In: 2020 IEEE International Conference on Artificial Intelligence and Virtual Reality (AIVR). p. 151–157. IEEE, Utrecht, Netherlands (2020)

12. Khamaj, A., Ali, A.M.: Adapting user experience with reinforcement learning: Personalizing interfaces based on user behavior analysis in real-time. Alexandria Engineering Journal 95, 164–173 (2024)

13. Quintero, L.: User Modeling for Adaptive Virtual Reality Experiences : Personalization from Behavioral and Physiological Time Series. Ph.D. thesis, Stockholm University, Department of Computer and Systems Sciences (2023)

14. Zhou, S., Zheng, W., Xu, Y., Liu, Y.: Enhancing user experience in vr environments through ai-driven adaptive ui design. Journal of Artificial Intelligence General science (JAIGS) ISSN:3006-4023 6(1), 59–82 (2024)

15. Saranya, S., Channarayapriya, B., Harshavardhini, U., Nandhini, A., Revathi, J., Venkatesan, R.: Development of virtual reality platform through human computer interaction using artificial intelligence. In: 2024 3rd International Conference on Applied Artificial Intelligence and Computing (ICAAIC). pp. 283–288 (2024)

16. Hadadi, A., Chardonnet, J.R., Guillet, C., Ovtcharova, J.: Smartsimvr: An architecture integrating machine learning and virtual environment for real-time simulation adaptation. In: 2024 10th International Conference on Automation, Robotics and Applications (ICARA). pp. 531–535. IEEE (2024)

17. Islam, R., Desai, K., Quarles, J.: Cybersickness prediction from integrated hmd's sensors: A multimodal deep fusion approach using eye-tracking and head-tracking data. In: 2021 IEEE International Symposium on Mixed and Augmented Reality (ISMAR). pp. 31–40 (2021)

18. Maneuvrier, A., Nguyen, N.D.T., Renaud, P.: Predicting vr cybersickness and its impact on visuomotor performance using head rotations and field (in)dependence. Frontiers in Virtual Reality 4, 1307925 (Nov 2023)

19. Breiman, L.: Random forests. Machine Learning 45(1), 5–32 (2001)

20. Unity Technologies: Unity game engine (2022), https://unity.com/, accessed 2024-03-12.

21. Python Software Foundation: Python language reference, version 3.11 (2022), https://www.python.org/, accessed 2024-03-12.

22. ONNX Community: ONNX: Open neural network exchange (2022), https://onnx.ai/, accessed 2024-03-12.

23. Unity Technologies: Unity barracuda: Neural network inference library (2022), https://docs.unity3d.com/Packages/com.unity.barracuda@1.0/manual/index.html, accessed 2024-03-12.

24. Kim, Y.H., Park, E.C., Choi, S.H., Choi, J.I.: Virtual reality sickness questionnaire (vrsq): Motion sickness measurement index in a virtual reality environment. Applied Ergonomics 69, 66–73 (2018)

25. Pedregosa, F., Varoquaux, G., Gramfort, A., Michel, V., Thirion, B., Grisel, O., Blondel, M., Prettenhofer, P., Weiss, R., Dubourg, V., Vanderplas, J., Passos, A., Cournapeau, D., Brucher, M., Perrot, M., Duchesnay, É.: Scikit-learn: Machine learning in python. Journal of Machine Learning Research 12, 2825–2830 (2011)

26. James, G., Witten, D., Hastie, T., Tibshirani, R.: An Introduction to Statistical Learning: With Applications in R. Springer Texts in Statistics, Springer (2013)